\documentstyle[12pt]{article}    
\begin{document}
\begin{center}
   
        {\bf  Quantum Energy of Fuzzy Sphere: Gaussian Variational Method}\footnote{Short summary of the Master-thesis}

\vspace{1cm}

                       Yu-Jie Zhong, Wung-Hong Huang\\
                       Department of Physics\\
                       National Cheng Kung University\\
                       Tainan,70101,Taiwan\\

\end{center}
\vspace{2cm}

    The  matrix model with mass term has a nontrivially classical solution which is known to represent a noncommutative fuzzy sphere.    The fuzzy sphere has a lower energy then that of the trivial solution.    In this letter we investigate the quantum correction of the energy of the fuzzy sphere by using  the Gaussian variational technique, in contrast to the other studying in which only the small fluctuation was considered.    Our result, which only considering the boson part, shows that the quantum correction does not change the stability of the fuzzy sphere. 

\vspace{3cm}
\begin{flushleft}
\end{flushleft}

\newpage

     The noncommutative fuzzy sphere, which can appear in the string/M theory [1-3], is known to correspond sphere D2-branes in string theory with background linear B-field [4].  It has been found that, in the presence of  constant RR three-form potential, the D0-branes can expand into a noncommutative fuzzy sphere configurations [5].   In the ordinary matrix model one could not find the fuzzy sphere solutions.   However, adding a Chern-Simons term [5] or mass term [6] into the matrix model will enable us to describe the noncommutative fuzzy sphere as a classical solution.   Comparing the energy in the various classical solutions one can find that the separated D0-branes will expand into a largest noncommutative fuzzy sphere to achieve minimum energy [4,5]. 

   In this paper we will study the matrix model with the  action 

$$ S = Tr  [X_i, X_j]^2  + \lambda ~Tr (X_i)^2. \eqno{(1)}$$
\\
where $X_i,  i=1,2,3$ are chosen to be $2 \times 2$ matrices for simplicity.  The matrix equations of (1) are
 
$$ 2  [X_j,  [X_i,X_j]] + \lambda  X_i = 0, \eqno{(2)}$$
\\
and the associated energy is 

$$ E =  - Tr  [X_i, X_j]^2  - \lambda Tr (X_i)^2.  \eqno{(3)} $$
\\
Eq.(2) admits commutating solutions and static fuzzy sphere solutions [4,5].   The commutating solutions are known to represent $N$ D0 branes and satisfy the relations
     $$[X_i,X_j] =0,   \eqno {(4)}$$
which  have the energy
     $$E= 0 .\eqno {(5)}$$ 
\\
The noncommutative static fuzzy sphere solutions satisfy the equations

$$[X_i, X_j] = i \sqrt {\lambda \over 2} \epsilon _{ijk} X_k,  \eqno {(6)}$$
\\
and can be described by the relations

   $$X_i = \sqrt {\lambda \over 2}~{\sigma_i \over 2},  \eqno{(7)}$$
\\
in which $\sigma_i$ is a Pauli spinor.  The energy E of  this solution has a negative energy

     $$E = - {\lambda \over 4} < 0 .\eqno {(8)}$$ 
\\
Thus the ground state is the 2-dimensional  fuzzy sphere [5].   

    In the following we will use the Gaussian variational method [7] to calculated the quantum energy of the above fuzzy sphere.   Note that, as a first step,  we consider the simplest $2\times 2$ matrix and only the mass term, rather then the conventional Chen-Simon term,  is presented in (1), as the calculation becomes more simpler.   We also consider only the boson part and leave the fermion part to the further investigations. 

    Following the method in [7] our trying Gaussian action is

$$ S_0 = \sum _{i= 1} ^{3}  \frac{3}{v} Tr ( X_i )^2,  \eqno{(9)}$$
\\
in which $v$ is a parameter to be determined by variation.   Then we can rewrite the partition function as 

   $$ Z   = Z_0  \langle  e ^{- (S-S_0)} \rangle _0  , \eqno{(10)}$$

    $$ Z_0 = \int \d X  e ^{-S_0},\eqno{(11)}$$ 
\\
where $\langle \ \cdot \ \rangle_0$ is a VEV with respect to the
partition function $Z_0$.   The associated free energy can be expanded as

$$ F = \sum_{k=0}^{\infty} F_k  ~~~;~~~F_0 = - ln Z_0 ,  \eqno{(12)}$$
$$ F_k =  -  \frac{(-1)^k}{k!} \langle (S - S_0)^k \rangle_{{\rm C},0}, \eqno{(13)}$$ 
\\
where the suffix `C' in $\langle \ \cdot \ \rangle _{{\rm C} , 0}$
means that the connected part is taken.

   After the calculations the first two terms of the expansion are given as

    $$F_0 = - \frac{9}{2}  \ln v , \eqno{(14)}$$ 
    $$F_1 =  \langle   S  \rangle_0 - \langle   S_0 \rangle_0 ,\eqno{(15)}$$ 
    $$ \langle   S  \rangle_0 = - \frac{9}{2} v^2 - \frac{3}{4}\lambda v, \eqno{(16)}$$
    $$\langle   S_0 \rangle_0 = \frac{9}{2}. \eqno{(17)}$$
\\
The variational parameters $v $ can be determined in such a way that the free energy $F$ becomes minimum, i.e. 

  $$ 0 = \frac{d}{d v} (F_0 + F_1).  \eqno{(18)}$$
\\
The solution is 

  $$ v = {-\lambda + \sqrt{\lambda^2+16}\over 8}, \eqno{(19)}$$
\\
and  free energy becomes

$$ F_0 + F_1 = - {3\over 2} ln \left(- \lambda +  \sqrt{\lambda^2+16}\right). \eqno{(20)}$$
\\
Therefore the quantum corrected energy coming from the boson part of the fuzzy sphere is negative and the value is  lower then that of the trivial solution.    Our result thus shows that the quantum correction does not change the stability of the fuzzy sphere.  
\\
\\

   Let us finally make some remarks about our works.

    (1)   As the fuzzy sphere is described by a supersymmetry action, a complete investigation shall contain the fermion part.  This part is under calculations.

   (2)   We consider only the simplest case of $2\times 2$ matrix, in stead of $N\times N$ matrix, as the calculation becomes more simpler.   A complete investigation shall study all possible represent of the SU(2) matrix.   Note that, from the one-loop action,  Iso etc. [8] had found that, in some situation, the fuzzy sphere  may becomes unstable in the large N limit.  It is remained to compare the final results for the one-loop perturbative method and Gaussian variational nonperturbative study.

  (3)   It is remained to compare the quantum properties of the fuzzy sphere coming from the Chern-Simon term with that coming from the mass term.    The stability of the fuzzy sphere at finite temperature is also an interesting problem to be studied.

   We propose to study these in the near future. \footnote{After we have completed this paper we find that the authors of the paper [9] also propose to study the problem.}  
\\
\\
\\
{REFERENCES}

\begin{enumerate}
\item  A. Connes, M. R. Douglas and A. Schwarz,
  ``Noncommutative Geometry and Matrix Theory: Compactification on
  Tori'', JHEP 9802:003 (1998), hep-th/9711162; \\
 B.~Morariu and B.~Zumino, ``Super Yang-Mills on the
  Noncommutative Torus'', hep-th/9807198; \\
 C.~Hofman and E.~Verlinde, ``U-duality of Born-Infeld on the
Noncommutative Two-Torus'', JHEP {\bf 9812}, 010 (1998),  hep-th/9810116 .
\item  N.~Seiberg and E.~Witten,
``String Theory and Noncommutative Geometry'', JHEP {\bf 9909}, 032 (1999),  hep-th/9908142.
\item  J. Polchinski, {\it String Theory}, Cambridge University Press,
1998.
\item  A.~Y.~Alekseev, A.~Recknagel and V.~Schomerus, ``Non-commutative world-volume geometries: Branes on su(2) and fuzzy sphere'' JHEP {\bf 9909} (1999) 023 [hep-th/9908040]; ``Brane dynamics in background fluxes and non-commutative geometry,'' JHEP {\bf 0005} (2000) 010 [hep-th/0003187].
\item  R.~C.~Myers, ``Dielectric-branes,'' JHEP {\bf 9912} (1999) 022 [hep-th/9910053]. 
\item  J. Polchinski and M. J. Strassler,  ``The String Dual of a Confining Four-Dimensional Gauge Theory'', [hep-th/0003136]. 
\item J. Nishimura and F. Sugino,  ``Dynamical Generation of Four-Dimensional\\
Space-Time in the IIB Matrix Model '', JHEP {\bf 0205} (2002) 001 [hep-th/0111102]. 
\item S. Iso, Y. Kimura, K. Tanaka, and K. Wakatsuki,  ``Noncommutative Gauge Theory on Fuzzy Sphere from Matrix Model '', Nucl. Phys. {\bf 604} (2001) 121 [hep-th/0101102]. 
\item T. Imai,  Y. Kitazaya, Y. Takayama, and D. Tomino,  ``Quantum Correction on Fuzzy Sphere'', [hep-th/0303120].

\end{enumerate}
\end{document}